%% file: nonlinear.tex
\documentclass[11pt]{article}
\usepackage{relsize}
\usepackage{amsmath}
\usepackage{amssymb}
\usepackage{xspace}
\usepackage[numbers]{natbib}
\usepackage[textwidth=480pt,textheight=680pt]{geometry}
\usepackage{graphicx}
\usepackage{color}
\usepackage[bottom]{footmisc}
\usepackage[bf,small]{caption2}
\usepackage{url}
\setcaptionwidth{.9\textwidth}
\makeatletter
\@addtoreset{equation}{section}

\makeatother

\DeclareMathOperator{\Sym}{Sym}

\providecommand{\href}[2]{#2}

\newdimen\tableauside\tableauside=1ex   
\newdimen\tableaurule\tableaurule=.32pt   
\newdimen\tableaustep
\def\phantomhrule#1{\hbox{\vbox to0pt{\hrule height\tableaurule width#1\vss}}}
\def\phantomvrule#1{\vbox{\hbox to0pt{\vrule width\tableaurule height#1\hss}}}
\def\sqr{\vbox{%
  \phantomhrule\tableaustep
  \hbox{\phantomvrule\tableaustep\kern\tableaustep\phantomvrule\tableaustep}%
  \hbox{\vbox{\phantomhrule\tableauside}\kern-\tableaurule}}}
\def\squares#1{\hbox{\count0=#1\noindent\loop\sqr
  \advance\count0 by-1 \ifnum\count0>0\repeat}}
\def\tableau#1{\vcenter{\offinterlineskip
  \tableaustep=\tableauside\advance\tableaustep by-\tableaurule
  \kern\normallineskip\hbox
    {\kern\normallineskip\vbox
      {\gettableau#1 0 }%
     \kern\normallineskip\kern\tableaurule}%
  \kern\normallineskip\kern\tableaurule}}
\def\gettableau#1 {\ifnum#1=0\let\next=\null\else
  \squares{#1}\let\next=\gettableau\fi\next}

\newcommand{\oplusr}{\raise2ex\hbox{$\oplus$}}
\newcommand{\otimesr}{\raise2ex\hbox{$\otimes$}}

\begin{document}
\pagestyle{empty}
\begin{flushright}
AEI-2005-113\\
DAMTP-2005-54\\
hep-th/0506161\\
June 20th, 2005
\end{flushright}
\vskip 5ex

\begin{center}
\begin{minipage}{.9\textwidth}
\lineskip 2ex
{\huge\bf Superfield integrals in high dimensions}\\[6ex]
{\bfseries Michael B.~Green}\\[1ex]
\hbox{~~~}\begin{minipage}{.8\textwidth}
Department of Applied Mathematics and Theoretical Physics\\
Centre for Mathematical Sciences\\
Wilberforce Road\\
CB3 0WA Cambridge, UK\\
{\tt M.B.Green@damtp.cam.ac.uk}
\end{minipage}\\[3ex]
{\bfseries Kasper Peeters}\\[1ex]
\hbox{~~~}\begin{minipage}{.8\textwidth}
Max-Planck-Institut f\"ur Gravitationsphysik\\
Albert-Einstein-Institut\\
Am M\"uhlenberg 1\\
14476 Golm, GERMANY\\
{\tt kasper.peeters@aei.mpg.de}
\end{minipage}\\[3ex]
{\bfseries Christian Stahn}\\[1ex]
\hbox{~~~}\begin{minipage}{.8\textwidth}
Department of Physics\\
University of North Carolina\\
Chapel Hill, NC 27599-3255\\
USA\\
{\tt stahn@physics.unc.edu}
\end{minipage}\\[7ex]

\noindent {\bf Abstract:} We present an efficient, covariant,
graph-based method to integrate superfields over fermionic spaces of
high dimensionality. We illustrate this method with the computation of
the most general sixteen-dimensional Majorana-Weyl integral in ten
dimensions. Our method has applications to the construction of
higher-derivative supergravity actions as well as the computation of
string and membrane vertex operator correlators.
\end{minipage}
\end{center}
\vfill
\newpage
\pagestyle{plain}
\hrule
\tableofcontents
\bigskip\medskip
\hrule
\bigskip

\section{Introduction}

Despite the conceptual elegance of superspace methods, their use in
the study of supergravity and string theory is hampered by a number of
technical difficulties. The problem which is perhaps most manifest is
the fact that high-dimensional fermionic integrals, although
conceptually simple, are hard to evaluate explicitly. This forms a
serious obstacle when one attempts to relate the often elegant
superspace expressions to results in terms of supergravity component
fields.

The fermionic integration problem is encountered in a variety of
different situations. One of these is the construction of
ten-dimensional higher-derivative supergravity actions. By virtue of
the presence of the dilaton field, these supergravity theories allow
for the existence of scalar superfields which contain the entire
supergravity multiplet in their component
expansion~\cite{nils1,howe1}. This observation has led (now almost
twenty years ago) to the hope that complicated higher-derivative
actions can perhaps be constructed in terms of superspace integrals of
simple expressions~\cite{nils2,Kallosh:1987mb}. Although the
application of this idea to the type-IIB theory is beset with
difficulties~\cite{deHaro:2002vk}, even the simpler construction
in~$N=1$ supergravity has never been worked out in full detail. One
particular problem which has remained unsolved is how to relate
superspace expressions to component ones. This same problem also
appears in the computation of vertex operator correlators for
superparticles~\cite{Green:1999by}, strings and
membranes~\cite{Dasgupta:2000df}, when these are formulated using
target-space spinors (i.e.~using the Green-Schwarz or Berkovits
formalisms). High-dimensional fermionic integrals appear here when one
tries to integrate out fermionic zero modes of the fields living on
the world-volume.

While generic covariant fermionic integrals are thus so far not known,
special cases which exhibit additional symmetries are sometimes
tractable. One such case is the integral that leads to the
sixteen-dilatino interaction in the type-IIB
theory~\cite{Green:1997me,Green:1998by}. The superspace supergravity
computation and the computation of a sixteen-fermion vertex operator
correlator both lead to the trivial integral
\begin{equation}
\label{e:lambda16}
\int\!{\rm d}^{16}\theta\, \big(\theta_a\lambda^a\big)^{16} = 
\epsilon_{a_1\cdots a_{16}} \lambda^{a_1}\cdots \lambda^{a_{16}} 
= 16!\,\lambda^1\cdots\lambda^{16}\,.
\end{equation}
A slightly more complicated expression is obtained for the
four-graviton amplitude in the Green-Schwarz formalism, when computed
in the light-cone gauge. The resulting fermionic integral can be
decomposed as the product of two known SO(8) fermionic
integrals~\cite{b_gree2}, with the result~\cite{Green:1997tv}
\begin{equation}
\label{e:graviton4}
\int\!{\rm d}^{8}\vartheta {\rm d}^8 \dot\vartheta
\Big((\bar{\vartheta}\gamma^{mn}\vartheta)(\bar{\dot{\vartheta}}\gamma^{rs}\dot{\vartheta})
R_{mn \, rs}\Big)^4 = \big( t_8 t_8 -
\tfrac{1}{4}\epsilon_8\epsilon_8\big) R^4\,.
\end{equation}
Again, this integral also occurs in an analysis of four-graviton
couplings in supergravity. A similar result can also be obtained for
the four-point correlator of supermembrane vertex operators in the
light-cone gauge~\cite{Dasgupta:2000df}. However, it is clear
that~\eqref{e:lambda16} and~\eqref{e:graviton4} form only the tip of
the iceberg. Many interesting results wait to be derived once a fully
covariant way is established to perform an arbitrary high-dimensional
fermionic superintegral and express it in terms of Lorentz singlets
(Kronecker deltas and epsilon tensors).

Therefore, it is the purpose of this small note to discuss a generic,
covariant method for the integration of arbitrary functions over
high-dimensional fermionic spaces. We demonstrate the feasibility of
our method by deriving an explicit expression for the most general
sixteen-component SO(10) fermionic integral (the ``$N=1$ integral'')
in terms of Lorentz singlets. This result is rather interesting by
itself, and we will discuss our motivation to derive it, including
possible applications, in some more detail towards the end (in
section~\ref{s:discuss}). The method can be applied easily to the
other ten- and eleven-dimensional supergravity theories.  Expressions
for SO(9) integrals, relevant for superparticle and supermembrane
calculations in eleven-dimensional supergravity in the light-cone
gauge, will appear shortly~\cite{plef1}.

For completeness, we also describe, in the appendix, an efficient
method to reduce tensor polynomials to a minimal basis. This simple
method does not seem to be widely known, but is of considerable help
in dealing with higher-derivative Lorentz invariants.

\section{Fermionic integrals}
\subsection{A simple eight-fermion example}
\label{s:t8}

The goal of this note is to show how high-dimensional fermionic
integrals can be evaluated covariantly and in full
generality. However, the techniques which we will use apply also to
much simpler cases. It is therefore illustrative to first consider a
simpler fermionic integral, which can be done by hand and for which
the answer has been known for a long time, so as to get familiar with
the techniques.

Let us thus consider the following integral over the eight-dimensional
space of~SO(8) spinors~\cite{b_gree2},
\begin{equation}
\label{e:It8}
I^{i_1j_1\cdots i_4j_4}_{\pm} := \int\!{\rm d}^8\theta^{\pm}\, 
     \big(\theta^{\pm} \gamma^{i_1 j_1} \theta^{\pm}\big)\cdots
     \big(\theta^{\pm} \gamma^{i_4 j_4} \theta^{\pm}\big)\,.
\end{equation}
Here the $\pm$ symbols denote the chirality of the spinors.  In order
to determine the number of Lorentz singlets which is needed to express
this integral, we compute the tensor product of the four symmetrised
sets of two anti-symmetric vector indices~\cite{e_cohe1},
\begin{equation}
\Sym^4\big([010\ldots]) = \left( \tableau{1 1} \right)^4_{\text{sym}}  
 = \begin{cases}
 3\times [0000] \oplus \ldots & \text{in SO(8)},\\[1ex]
 2\times [00000\ldots] \oplus \ldots & \text{in SO(2$k$) for $k>4$}.
\end{cases}
\end{equation}
This result implies that~\eqref{e:It8} can be decomposed in two delta
singlets and one epsilon singlet (the epsilon singlet is dimension
dependent and corresponds to the disappearing singlet when the tensor
product is evaluated in higher dimensions). It is straightforward to
find these three independent singlets; we will use
\begin{equation}
\begin{aligned}
D_1 &= \frac{1}{12}\big( \delta^{i_1 i_2} \delta^{j_1 j_2} 
                         \delta^{i_3 i_4} \delta^{j_3 j_4} + \text{11 terms}\big)\,,\\[1ex]
D_2 &= \frac{1}{48}\big( \delta^{j_1 i_2} \delta^{j_2 i_3} 
                         \delta^{j_3 i_4} \delta^{j_4 i_1} + \text{47 terms}\big)\,,\\[1ex]
E   &= \varepsilon^{i_1 j_1 i_2\cdots j_4}\,.
\end{aligned}
\end{equation}
The fermionic integral can thus be written as
\begin{equation}
I^{i_1j_1\cdots i_4j_4}_{\pm} = \alpha_1\, D^{i_1j_1\cdots i_4j_4}_1
 + \alpha_2\, D^{i_1j_1\cdots i_4j_4}_2
 \pm \beta\, E^{i_1 j_1 i_2\cdots j_4}\,,
\end{equation}
and the goal is to determine the unknown coefficients~$\alpha_1,
\alpha_2$ and $\beta$.

By using an explicit representation for the~SO(8) gamma matrices, it
is straightforward to evaluate~\eqref{e:It8} for particular values of
the eight indices (by selecting the terms in the resulting polynomial
of spinor components in which each component occurs once). Similarly,
it is straightforward to determine the value of $D_1$, $D_2$ and $E$
for a particular set of index values. Three independent combinations
are listed below,
\begin{equation}
\begin{tabular}{c|c|c|c|c|c}
{}[$i_1 j_1$]$\cdots$ [$i_4 j_4$] & $I_+$ & $I_-$ & $D_1$ & $D_2$ & $E$ \\
\hline
{}[12][12][34][34] & $-128$ & $-128$ & 1/12 & 0 & 0 \\
{}[12][23][34][41] & $\phantom{-}128$  & $\phantom{-}128$  & 0 & 1/48 & 0 \\
{}[12][34][56][78] & $\phantom{-}128$  & $-128$ & 0 & 0 & 1 
\end{tabular}
\end{equation}
This leads to three equations for three unknowns, from which one
determines the coefficients to be
\begin{equation}
\alpha_1 = -1536\,,\quad
\alpha_2 = 6144\,,\quad
\beta = 128\,.
\end{equation}
Comparing with the $t_8$ tensor of appendix~9.A of~\cite{b_gree2}, one
then obtains the expected result
\begin{equation}
I_{\pm} = 256\,t_{8,\pm}\,.
\end{equation}

In the next section we will see that the sixteen fermion integral can
be evaluated using precisely the same logic, although the number of
Lorentz singlets increases sharply and it also becomes more
complicated to evaluate their values given a set of indices. This
increased complexity calls for a number of new ideas.

\subsection{The sixteen fermion integral}

Let us now turn to the evaluation of the sixteen fermion integral in
ten-dimensional simple supergravity. The Weyl and Majorana properties
of the spinor~$\theta$ imply that bilinears in~$\theta$ can be written
as three-forms. The most general integrand therefore has the form
\begin{equation}
I^{i_1j_1k_1\cdots i_8 j_8 k_8} :=
\label{e:theintegral}
\int\!{\rm d}^{16}\theta\,
(\bar\theta\Gamma^{i_1 j_1 k_1}\theta)
(\bar\theta\Gamma^{i_2 j_2 k_2}\theta)\cdots
(\bar\theta\Gamma^{i_8 j_8 k_8}\theta) \,.
\end{equation}
The goal will again be to express this integral in terms of Lorentz
singlets, i.e.~Kronecker deltas and epsilon tensors carrying the free
vector indices. That is, we want to write the integral as
\begin{equation}
\label{e:decomposition}
I^{i_1j_1k_1\cdots i_8 j_8 k_8} =
 \sum_{i} \alpha_{i}\, T_{(i)}^{i_1j_1k_1\cdots i_8 j_8 k_8}\,.
\end{equation} 
in terms of a set of basis tensors~$T_{(i)}$.

The first step in this integration is to determine the number of
Lorentz singlets in which the integral~\eqref{e:theintegral} can be
decomposed. This number is easily obtained by considering the tensor
product
\begin{equation}
\label{e:sym8alt3}
\Sym^8\big(\,[00100\ldots]\,\big) = \left( \tableau{1 1 1} \right)^8_{\text{sym}}  
 = 
\begin{cases}
 33\times [00000] \oplus \ldots & \text{in SO(10),}\\[1ex]
 24\times [000\ldots] \oplus \ldots & \text{in SO($2k$) for $k>5$.}
\end{cases}
\end{equation}
There are thus 33 singlets in SO(10). Nine of these disappear when one
considers higher-dimensional spaces, and these thus correspond to
parity-odd invariants involving the ten-dimensional epsilon tensor.

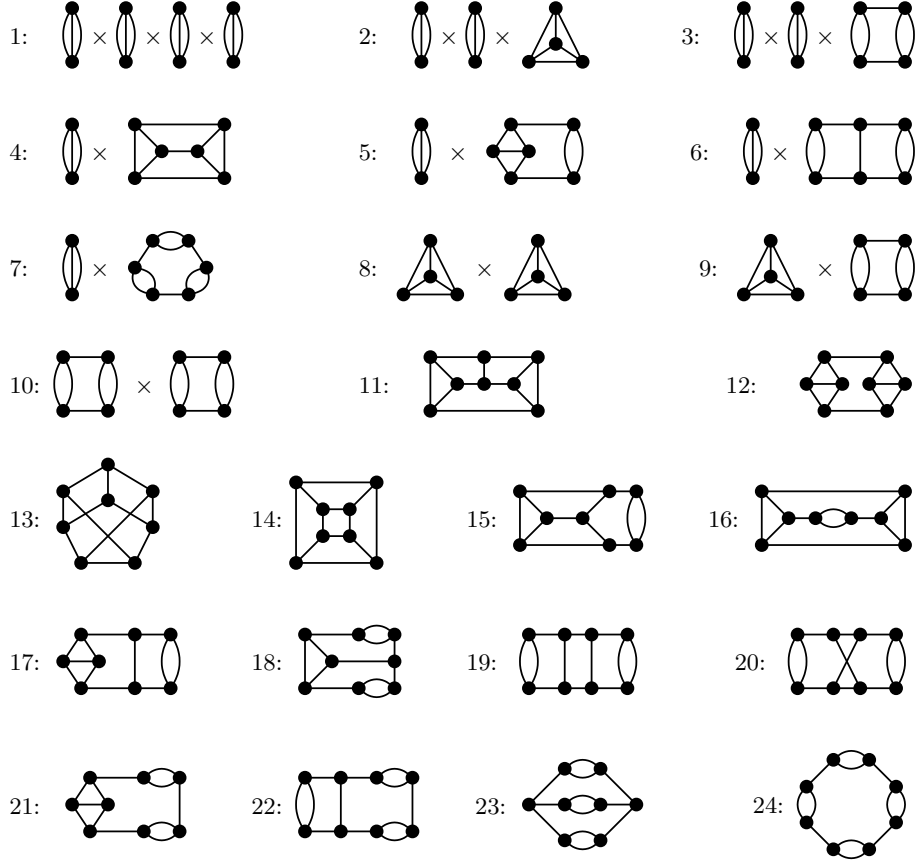
\begin{figure}[t]
\begin{center}
\input{graphs24.pstex_t}
\caption{Graphical display of the 24 Lorentz singlets in the
  $\theta^{16}$ integral which can be expressed solely using Kronecker
  deltas. Each dot represents a fully anti-symmetrised index triplet
  (e.g.~$[i_1,j_1,k_1]$) and the lines indicate how these indices
  appear on Kronecker delta symbols.}
\label{fig:graphs24}
\end{center}
\end{figure}

\begin{figure}[t]
\begin{center}
\input{zerographs.pstex_t}
\caption{Graphs of parity-even type which vanish identically by virtue
of the identities~\eqref{e:vanish1} and~\eqref{e:vanish2}.}
\label{fig:vanish}
\end{center}
\end{figure}
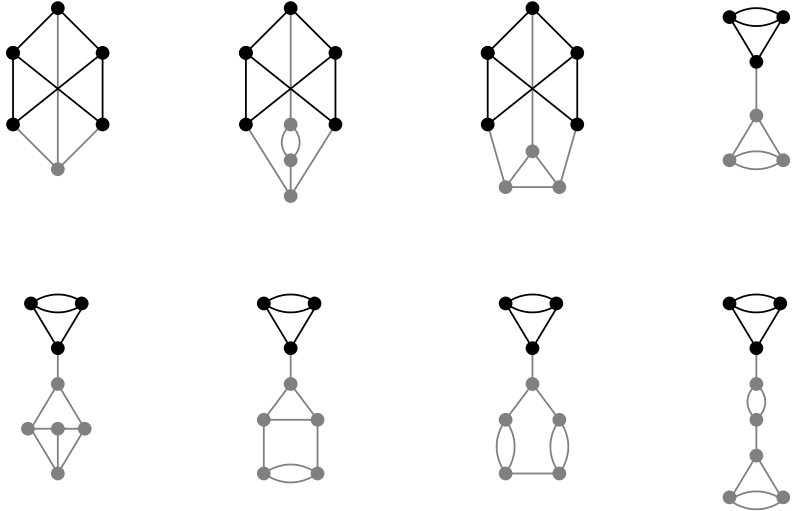

In contrast to the situation in the previous section, it is now not so
easy to guess the 33~Lorentz singlets. However, the explicit
construction of these singlets can be translated to an elegant problem
in graph theory. Let us first focus on the singlets which involve only
Kronecker deltas. We will represent fully anti-symmetrised index
triplets by trivalent nodes, and represent Kronecker deltas which set
two indices equal by edges connecting the nodes. Multiple edges are
allowed (e.g.~$\delta^{i_1 i_2} \delta^{j_1 j_2} \delta^{k_1 k_2}$,
which corresponds to two nodes with a 3-fold connection between
them). The problem of finding all~24 independent singlets now
corresponds, in standard graph terminology, to the problem of finding
all 3-regular multigraphs (not necessarily connected) with eight
vertices.  All graphs of this type can be found with the help
of~\cite{atlas,e_meringer}, and one finds a total of 32 graphs. This
may seem to contradict~\eqref{e:sym8alt3}, however, the anti-symmetry
of the index triplets is not yet fully encoded at this stage. Using
anti-symmetry,~8 of the 32 graphs can be shown to correspond to a
vanishing expression. For example, the double contraction of two
identical three-forms is symmetric in the two free indices and
therefore vanishes when contracted into a further three-form. In
graphical notation, this is seen from
\begin{equation}
\label{e:vanish1}
\input{proofzero11.pstex_t}
\end{equation}
Another vanishing subgraph is
\begin{equation}
\label{e:vanish2}
\input{proofzero21.pstex_t}
\end{equation}
These identities make 8 graphs vanish identically, see
figure~\ref{fig:vanish}. The remaining 24 graphs are listed in
figure~\ref{fig:graphs24} and their explicit expressions in terms of
Kronecker deltas can be found in table~\ref{tab:deltaresults}.

\begin{table}[p]
\centering
\begin{tabular}{c|c|c}
Graph $i$ & Singlet $T_{(i)}$ & Coefficient \\
          &                   & $\alpha_i / (2^{19} \cdot 3^6)$ \\
\hline
&& \\[-3pt]
1 & $\delta_{{i_2}}^{{i_1}}\,\delta_{{i_4}}^{{i_3}}\,\delta_{{i_6}}^{{i_5}}\,\delta_{{i_8}}^{{i_7}}\,\delta_{{j_2}}^{{j_1}}\,\delta_{{j_4}}^{{j_3}}\,\delta_{{j_6}}^{{j_5}}\,\delta_{{j_8}}^{{j_7}}\,\delta_{{k_2}}^{{k_1}}\,\delta_{{k_4}}^{{k_3}}\,\delta_{{k_6}}^{{k_5}}\,\delta_{{k_8}}^{{k_7}}$ & $-269$ \\[3pt]
2 & $\delta_{{i_2}}^{{i_1}}\,\delta_{{i_4}}^{{i_3}}\,\delta_{{i_6}}^{{i_5}}\,\delta_{{j_6}}^{{i_7}}\,\delta_{{k_5}}^{{i_8}}\,\delta_{{j_2}}^{{j_1}}\,\delta_{{j_4}}^{{j_3}}\,\delta_{{j_7}}^{{j_5}}\,\delta_{{k_6}}^{{j_8}}\,\delta_{{k_2}}^{{k_1}}\,\delta_{{k_4}}^{{k_3}}\,\delta_{{k_8}}^{{k_7}}$ & $4968$ \\[3pt]
3 & $\delta_{{i_2}}^{{i_1}}\,\delta_{{i_4}}^{{i_3}}\,\delta_{{i_6}}^{{i_5}}\,\delta_{{k_6}}^{{i_7}}\,\delta_{{k_5}}^{{i_8}}\,\delta_{{j_2}}^{{j_1}}\,\delta_{{j_4}}^{{j_3}}\,\delta_{{j_6}}^{{j_5}}\,\delta_{{j_8}}^{{j_7}}\,\delta_{{k_2}}^{{k_1}}\,\delta_{{k_4}}^{{k_3}}\,\delta_{{k_8}}^{{k_7}}$ & $7956$ \\[3pt]
4 & $\delta_{{i_2}}^{{i_1}}\,\delta_{{i_4}}^{{i_3}}\,\delta_{{j_4}}^{{i_5}}\,\delta_{{k_3}}^{{i_6}}\,\delta_{{j_3}}^{{i_7}}\,\delta_{{k_5}}^{{i_8}}\,\delta_{{j_2}}^{{j_1}}\,\delta_{{j_6}}^{{j_5}}\,\delta_{{k_4}}^{{j_7}}\,\delta_{{k_6}}^{{j_8}}\,\delta_{{k_2}}^{{k_1}}\,\delta_{{k_8}}^{{k_7}}$ & $-2304$ \\[3pt]
5 & $\delta_{{i_2}}^{{i_1}}\,\delta_{{i_4}}^{{i_3}}\,\delta_{{j_3}}^{{i_5}}\,\delta_{{k_4}}^{{i_6}}\,\delta_{{k_3}}^{{i_7}}\,\delta_{{k_6}}^{{i_8}}\,\delta_{{j_2}}^{{j_1}}\,\delta_{{j_5}}^{{j_4}}\,\delta_{{k_5}}^{{j_6}}\,\delta_{{j_8}}^{{j_7}}\,\delta_{{k_2}}^{{k_1}}\,\delta_{{k_8}}^{{k_7}}$ & $70848$ \\[3pt]
6 & $\delta_{{i_2}}^{{i_1}}\,\delta_{{i_4}}^{{i_3}}\,\delta_{{k_3}}^{{i_5}}\,\delta_{{k_4}}^{{i_6}}\,\delta_{{k_5}}^{{i_7}}\,\delta_{{k_6}}^{{i_8}}\,\delta_{{j_2}}^{{j_1}}\,\delta_{{j_4}}^{{j_3}}\,\delta_{{j_6}}^{{j_5}}\,\delta_{{j_8}}^{{j_7}}\,\delta_{{k_2}}^{{k_1}}\,\delta_{{k_8}}^{{k_7}}$ & $-24192$ \\[3pt]
7 & $\delta_{{i_2}}^{{i_1}}\,\delta_{{i_4}}^{{i_3}}\,\delta_{{k_4}}^{{i_5}}\,\delta_{{j_5}}^{{i_6}}\,\delta_{{k_6}}^{{i_7}}\,\delta_{{j_7}}^{{i_8}}\,\delta_{{j_2}}^{{j_1}}\,\delta_{{j_4}}^{{j_3}}\,\delta_{{k_5}}^{{j_6}}\,\delta_{{k_7}}^{{j_8}}\,\delta_{{k_2}}^{{k_1}}\,\delta_{{k_8}}^{{k_3}}$ & $-32544$ \\[3pt]
8 & $\delta_{{i_2}}^{{i_1}}\,\delta_{{j_2}}^{{i_3}}\,\delta_{{k_1}}^{{i_4}}\,\delta_{{i_6}}^{{i_5}}\,\delta_{{j_6}}^{{i_7}}\,\delta_{{k_5}}^{{i_8}}\,\delta_{{j_3}}^{{j_1}}\,\delta_{{k_2}}^{{j_4}}\,\delta_{{j_7}}^{{j_5}}\,\delta_{{k_6}}^{{j_8}}\,\delta_{{k_4}}^{{k_3}}\,\delta_{{k_8}}^{{k_7}}$ & $-3888$ \\[3pt]
9 & $\delta_{{i_2}}^{{i_1}}\,\delta_{{j_2}}^{{i_3}}\,\delta_{{k_1}}^{{i_4}}\,\delta_{{i_6}}^{{i_5}}\,\delta_{{k_6}}^{{i_7}}\,\delta_{{k_5}}^{{i_8}}\,\delta_{{j_3}}^{{j_1}}\,\delta_{{k_2}}^{{j_4}}\,\delta_{{j_6}}^{{j_5}}\,\delta_{{j_8}}^{{j_7}}\,\delta_{{k_4}}^{{k_3}}\,\delta_{{k_8}}^{{k_7}}$ & $-26352$ \\[3pt]
10 & $\delta_{{i_2}}^{{i_1}}\,\delta_{{k_2}}^{{i_3}}\,\delta_{{k_1}}^{{i_4}}\,\delta_{{i_6}}^{{i_5}}\,\delta_{{k_6}}^{{i_7}}\,\delta_{{k_5}}^{{i_8}}\,\delta_{{j_2}}^{{j_1}}\,\delta_{{j_4}}^{{j_3}}\,\delta_{{j_6}}^{{j_5}}\,\delta_{{j_8}}^{{j_7}}\,\delta_{{k_4}}^{{k_3}}\,\delta_{{k_8}}^{{k_7}}$ & $-20412$ \\[3pt]
11 & $\delta_{{i_2}}^{{i_1}}\,\delta_{{j_1}}^{{i_3}}\,\delta_{{k_3}}^{{i_4}}\,\delta_{{k_1}}^{{i_5}}\,\delta_{{k_4}}^{{i_6}}\,\delta_{{k_2}}^{{i_7}}\,\delta_{{k_5}}^{{i_8}}\,\delta_{{j_3}}^{{j_2}}\,\delta_{{j_5}}^{{j_4}}\,\delta_{{j_8}}^{{j_6}}\,\delta_{{k_6}}^{{j_7}}\,\delta_{{k_8}}^{{k_7}}$ & $124416$ \\[3pt]
12 & $\delta_{{i_2}}^{{i_1}}\,\delta_{{j_1}}^{{i_3}}\,\delta_{{k_2}}^{{i_4}}\,\delta_{{k_1}}^{{i_5}}\,\delta_{{j_5}}^{{i_6}}\,\delta_{{k_5}}^{{i_7}}\,\delta_{{k_4}}^{{i_8}}\,\delta_{{j_3}}^{{j_2}}\,\delta_{{k_3}}^{{j_4}}\,\delta_{{j_7}}^{{j_6}}\,\delta_{{k_6}}^{{j_8}}\,\delta_{{k_8}}^{{k_7}}$ & $10368$ \\[3pt]
13 & $\delta_{{i_2}}^{{i_1}}\,\delta_{{j_1}}^{{i_3}}\,\delta_{{k_1}}^{{i_4}}\,\delta_{{j_2}}^{{i_5}}\,\delta_{{k_4}}^{{i_6}}\,\delta_{{k_3}}^{{i_7}}\,\delta_{{k_2}}^{{i_8}}\,\delta_{{j_6}}^{{j_3}}\,\delta_{{j_5}}^{{j_4}}\,\delta_{{k_5}}^{{j_7}}\,\delta_{{k_6}}^{{j_8}}\,\delta_{{k_8}}^{{k_7}}$ & $196992$ \\[3pt]
14 & $\delta_{{i_2}}^{{i_1}}\,\delta_{{j_2}}^{{i_3}}\,\delta_{{j_3}}^{{i_4}}\,\delta_{{k_1}}^{{i_5}}\,\delta_{{k_2}}^{{i_6}}\,\delta_{{k_3}}^{{i_7}}\,\delta_{{k_4}}^{{i_8}}\,\delta_{{j_4}}^{{j_1}}\,\delta_{{j_6}}^{{j_5}}\,\delta_{{k_6}}^{{j_7}}\,\delta_{{k_7}}^{{j_8}}\,\delta_{{k_8}}^{{k_5}}$ & $-10368$ \\[3pt]
15 & $\delta_{{i_2}}^{{i_1}}\,\delta_{{j_1}}^{{i_3}}\,\delta_{{k_3}}^{{i_4}}\,\delta_{{k_2}}^{{i_5}}\,\delta_{{k_1}}^{{i_6}}\,\delta_{{k_6}}^{{i_7}}\,\delta_{{k_5}}^{{i_8}}\,\delta_{{j_3}}^{{j_2}}\,\delta_{{j_5}}^{{j_4}}\,\delta_{{k_4}}^{{j_6}}\,\delta_{{j_8}}^{{j_7}}\,\delta_{{k_8}}^{{k_7}}$ & $373248$ \\[3pt]
16 & $\delta_{{i_2}}^{{i_1}}\,\delta_{{j_1}}^{{i_3}}\,\delta_{{k_3}}^{{i_4}}\,\delta_{{j_4}}^{{i_5}}\,\delta_{{k_5}}^{{i_6}}\,\delta_{{k_2}}^{{i_7}}\,\delta_{{k_1}}^{{i_8}}\,\delta_{{j_3}}^{{j_2}}\,\delta_{{k_4}}^{{j_5}}\,\delta_{{j_7}}^{{j_6}}\,\delta_{{k_6}}^{{j_8}}\,\delta_{{k_8}}^{{k_7}}$ & $-331776$\\[3pt]
17 & $\delta_{{i_2}}^{{i_1}}\,\delta_{{j_1}}^{{i_3}}\,\delta_{{k_2}}^{{i_4}}\,\delta_{{k_4}}^{{i_5}}\,\delta_{{k_1}}^{{i_6}}\,\delta_{{k_5}}^{{i_7}}\,\delta_{{k_6}}^{{i_8}}\,\delta_{{j_3}}^{{j_2}}\,\delta_{{k_3}}^{{j_4}}\,\delta_{{j_6}}^{{j_5}}\,\delta_{{j_8}}^{{j_7}}\,\delta_{{k_8}}^{{k_7}}$ & $-165888$ \\[3pt]
18 & $\delta_{{i_2}}^{{i_1}}\,\delta_{{j_1}}^{{i_3}}\,\delta_{{j_2}}^{{i_4}}\,\delta_{{j_3}}^{{i_5}}\,\delta_{{k_1}}^{{i_6}}\,\delta_{{k_4}}^{{i_7}}\,\delta_{{k_5}}^{{i_8}}\,\delta_{{k_2}}^{{j_4}}\,\delta_{{k_3}}^{{j_5}}\,\delta_{{j_7}}^{{j_6}}\,\delta_{{k_6}}^{{j_8}}\,\delta_{{k_8}}^{{k_7}}$ & $41472$ \\[3pt]
19 & $\delta_{{i_2}}^{{i_1}}\,\delta_{{k_1}}^{{i_3}}\,\delta_{{k_2}}^{{i_4}}\,\delta_{{k_3}}^{{i_5}}\,\delta_{{k_4}}^{{i_6}}\,\delta_{{k_5}}^{{i_7}}\,\delta_{{k_6}}^{{i_8}}\,\delta_{{j_2}}^{{j_1}}\,\delta_{{j_4}}^{{j_3}}\,\delta_{{j_6}}^{{j_5}}\,\delta_{{j_8}}^{{j_7}}\,\delta_{{k_8}}^{{k_7}}$ & $-10368$ \\[3pt]
20 & $\delta_{{i_2}}^{{i_1}}\,\delta_{{k_1}}^{{i_3}}\,\delta_{{k_2}}^{{i_4}}\,\delta_{{k_3}}^{{i_5}}\,\delta_{{j_3}}^{{i_6}}\,\delta_{{k_5}}^{{i_7}}\,\delta_{{k_6}}^{{i_8}}\,\delta_{{j_2}}^{{j_1}}\,\delta_{{j_6}}^{{j_4}}\,\delta_{{k_4}}^{{j_5}}\,\delta_{{j_8}}^{{j_7}}\,\delta_{{k_8}}^{{k_7}}$ & $-171072$ \\[3pt]
21 & $\delta_{{i_2}}^{{i_1}}\,\delta_{{j_1}}^{{i_3}}\,\delta_{{k_2}}^{{i_4}}\,\delta_{{k_1}}^{{i_5}}\,\delta_{{j_5}}^{{i_6}}\,\delta_{{k_6}}^{{i_7}}\,\delta_{{k_4}}^{{i_8}}\,\delta_{{j_3}}^{{j_2}}\,\delta_{{k_3}}^{{j_4}}\,\delta_{{k_5}}^{{j_6}}\,\delta_{{j_8}}^{{j_7}}\,\delta_{{k_8}}^{{k_7}}$ & $-238464$ \\[3pt]
22 & $\delta_{{i_2}}^{{i_1}}\,\delta_{{k_1}}^{{i_3}}\,\delta_{{k_2}}^{{i_4}}\,\delta_{{k_3}}^{{i_5}}\,\delta_{{k_4}}^{{i_6}}\,\delta_{{j_5}}^{{i_7}}\,\delta_{{k_7}}^{{i_8}}\,\delta_{{j_2}}^{{j_1}}\,\delta_{{j_4}}^{{j_3}}\,\delta_{{j_8}}^{{j_6}}\,\delta_{{k_5}}^{{j_7}}\,\delta_{{k_8}}^{{k_6}}$ & $248832$ \\[3pt]
23 & $\delta_{{i_2}}^{{i_1}}\,\delta_{{j_1}}^{{i_3}}\,\delta_{{k_1}}^{{i_4}}\,\delta_{{i_8}}^{{i_5}}\,\delta_{{j_8}}^{{i_6}}\,\delta_{{k_8}}^{{i_7}}\,\delta_{{j_5}}^{{j_2}}\,\delta_{{j_6}}^{{j_3}}\,\delta_{{j_7}}^{{j_4}}\,\delta_{{k_5}}^{{k_2}}\,\delta_{{k_6}}^{{k_3}}\,\delta_{{k_7}}^{{k_4}}$ & $-62208$\\[3pt]
24 & $\delta_{{i_2}}^{{i_1}}\,\delta_{{k_2}}^{{i_3}}\,\delta_{{j_3}}^{{i_4}}\,\delta_{{k_4}}^{{i_5}}\,\delta_{{j_5}}^{{i_6}}\,\delta_{{k_6}}^{{i_7}}\,\delta_{{j_7}}^{{i_8}}\,\delta_{{j_2}}^{{j_1}}\,\delta_{{k_3}}^{{j_4}}\,\delta_{{k_5}}^{{j_6}}\,\delta_{{k_7}}^{{j_8}}\,\delta_{{k_8}}^{{k_1}}$ & $63504$
\end{tabular}
\caption{The coefficients of the 24 Lorentz singlets in the
  $\theta^{16}$ integral which can be expressed solely using Kronecker
  deltas.\label{tab:deltaresults}  The singlets are understood to be antisymmetrised in the $[ijk]$ triplets, and symmetrised over $1\dots 8$.}
\end{table}

\begin{figure}
\begin{center}
\input{epsgraphs.pstex_t}
\caption{Graphical display of the 9 Lorentz singlets in the
  $\theta^{16}$ integral which contain epsilon tensors. Rectangles
  denote epsilon tensors, and each dot represents a fully
  anti-symmetrised index triplet, as in figure~\ref{fig:graphs24}.\label{fig:epsgraphs}}
\end{center}
\end{figure}
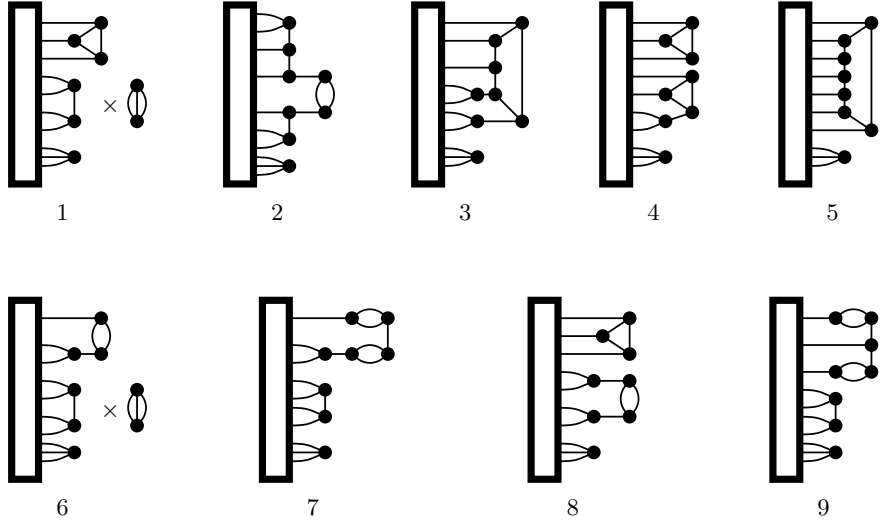

The construction of the graphs corresponding to parity-odd singlets is
less systematic. We can again introduce a graphical notation,
representing an epsilon tensor by a 10-valent vertex (or ``box").
Many graphs can be formed out of one 10-valent and eight trivalent
vertices, but nine of these are sufficient to represent the nine
independent parity-odd singlets in the tensor product
\eqref{e:sym8alt3}. These are displayed in figure~\ref{fig:epsgraphs}
and the corresponding explicit expressions can be found in
table~\ref{tab:epsresults}.

\begin{table}
\vspace{1ex}
\centering
\begin{tabular}{c|c|c}
Graph & Singlet & Coefficient \\
$i$   & $T_{(24+i)}$ & $\alpha_{24+i}/(2^{21} \cdot 3^6 \cdot5$) \\
\hline
&& \\[-3pt]
1 & ${{\eta }_{{i_1}{i_2}}}\,{{\eta }_{{i_3}{i_4}}}\,{{\eta}_{{i_5}{j_3}}}\,{{\eta }_{{i_6}{i_7}}}\,{{\eta }_{{j_1}{j_2}}}\,{{\eta}_{{j_4}{j_5}}}\,{{\eta }_{{k_1}{k_2}}}\,{{\epsilon}_{{i_8}{j_8}{k_8}{j_7}{k_7}{j_6}{k_6}{k_5}{k_4}{k_3}}}$ & $7$ \\[3pt]
2 & ${{\eta }_{{i_1}{i_2}}}\,{{\eta }_{{i_3}{k_1}}}\,{{\eta}_{{i_4}{k_2}}}\,{{\eta }_{{i_5}{j_3}}}\,{{\eta }_{{i_6}{j_4}}}\,{{\eta}_{{i_7}{j_5}}}\,{{\eta }_{{j_1}{j_2}}}\,{{\epsilon}_{{i_8}{j_8}{k_8}{j_7}{k_7}{j_6}{k_6}{k_5}{k_4}{k_3}}}$ & $42$ \\[3pt]
3 & ${{\eta }_{{i_1}{i_2}}}\,{{\eta }_{{i_3}{j_1}}}\,{{\eta}_{{i_4}{j_2}}}\,{{\eta }_{{i_5}{j_3}}}\,{{\eta }_{{i_6}{k_1}}}\,{{\eta}_{{i_7}{k_2}}}\,{{\eta }_{{j_4}{j_5}}}\,{{\epsilon}_{{i_8}{j_8}{k_8}{j_7}{k_7}{j_6}{k_6}{k_5}{k_4}{k_3}}}$ & $-294$ \\[3pt]
4 & ${{\eta }_{{i_1}{i_2}}}\,{{\eta }_{{i_3}{j_1}}}\,{{\eta}_{{i_4}{i_5}}}\,{{\eta }_{{i_6}{j_4}}}\,{{\eta }_{{i_7}{k_1}}}\,{{\eta}_{{j_2}{j_3}}}\,{{\eta }_{{j_5}{j_6}}}\,{{\epsilon}_{{i_8}{j_8}{k_8}{j_7}{k_7}{k_6}{k_5}{k_4}{k_3}{k_2}}}$ & $-168$ \\[3pt]
5 & ${{\eta }_{{i_1}{i_2}}}\,{{\eta }_{{i_3}{j_1}}}\,{{\eta}_{{i_4}{j_2}}}\,{{\eta }_{{i_5}{j_3}}}\,{{\eta }_{{i_6}{j_4}}}\,{{\eta}_{{i_7}{j_5}}}\,{{\eta }_{{j_6}{j_7}}}\,{{\epsilon}_{{i_8}{j_8}{k_8}{k_7}{k_6}{k_5}{k_4}{k_3}{k_2}{k_1}}}$ & $264$ \\[3pt]
6 & ${{\eta }_{{i_1}{i_2}}}\,{{\eta }_{{i_3}{i_4}}}\,{{\eta}_{{i_5}{k_3}}}\,{{\eta }_{{i_6}{i_7}}}\,{{\eta }_{{j_1}{j_2}}}\,{{\eta}_{{j_3}{j_4}}}\,{{\eta }_{{k_1}{k_2}}}\,{{\epsilon}_{{i_8}{j_8}{k_8}{j_7}{k_7}{j_6}{k_6}{j_5}{k_5}{k_4}}}$ & $0$ \\[3pt]
7 & ${{\eta }_{{i_1}{i_2}}}\,{{\eta }_{{i_3}{k_1}}}\,{{\eta}_{{i_4}{j_3}}}\,{{\eta }_{{i_5}{k_2}}}\,{{\eta }_{{i_6}{i_7}}}\,{{\eta}_{{j_1}{j_2}}}\,{{\eta }_{{j_4}{k_3}}}\,{{\epsilon}_{{i_8}{j_8}{k_8}{j_7}{k_7}{j_6}{k_6}{j_5}{k_5}{k_4}}}$ & $0$ \\[3pt]
8 & ${{\eta }_{{i_1}{i_2}}}\,{{\eta }_{{i_3}{i_4}}}\,{{\eta}_{{i_5}{j_3}}}\,{{\eta }_{{i_6}{k_1}}}\,{{\eta }_{{i_7}{k_2}}}\,{{\eta}_{{j_1}{j_2}}}\,{{\eta }_{{j_4}{j_5}}}\,{{\epsilon}_{{i_8}{j_8}{k_8}{j_7}{k_7}{j_6}{k_6}{k_5}{k_4}{k_3}}}$ & $0$ \\[3pt]
9 & ${{\eta }_{{i_1}{i_3}}}\,{{\eta }_{{i_2}{j_4}}}\,{{\eta}_{{i_4}{k_1}}}\,{{\eta }_{{i_5}{j_2}}}\,{{\eta }_{{i_6}{i_7}}}\,{{\eta}_{{j_1}{j_3}}}\,{{\eta }_{{j_5}{k_2}}}\,{{\epsilon}_{{i_8}{j_8}{k_8}{j_7}{k_7}{j_6}{k_6}{k_5}{k_4}{k_3}}}$ & $0$
\end{tabular}
\caption{Coefficients of the nine selected epsilon
  singlets\label{tab:epsresults} which occur in the fermionic integral~\eqref{e:theintegral}.  The vanishing of four of the coefficients has been achieved by choosing a suitable basis of the nine parity-odd singlets in~\eqref{e:sym8alt3}.}
\end{table}

Having determined the basis on which~\eqref{e:theintegral} can be
decomposed, the remaining step is to determine the coefficients in
front of each of these basis tensors. This is again done by matching
the values of~\eqref{e:theintegral} and~\eqref{e:decomposition} for
various sets of values of the indices, and solving the resulting
system of linear equations.\footnote{The chirality of the spinor
$\theta$ in~\eqref{e:theintegral} enters at this stage.  As in
section~\ref{s:t8}, the choice of chirality reflects itself in the
sign of the coefficients~\mbox{$\alpha_{25}\dots \alpha_{33}$} of the
parity-odd singlets, and we will not comment on this any further.}  In
graph-theory language, the evaluation of~\eqref{e:decomposition}
corresponds to finding all ways of colouring the edges of a graph with
numbers, given a set of three numbers at each vertex. This requires
some care in order to keep the computation within bounds. The basis
tensors in table~\ref{tab:deltaresults} and~\ref{tab:epsresults}
contain an implicit anti-symmetrisation over all indices in each
triplet, as well as an implicit symmetrisation over all index
triplets.  A brute-force algorithm which investigates the value of a
given singlet for all terms in the symmetrisation therefore leads to a
worst-case situation in which~$(3!)^8\cdot 8! \sim 6.8\times 10^{10}$
different terms (or colourings) have to be considered. The graphical
representation suggests a much more efficient ``backtracking''
algorithm. This algorithm constructs the graph labellings vertex by
vertex and checks after each choice if the index assignment is still
consistent. If it is, the next vertex is labelled, if not, the
algorithm backtracks and proceeds to the next choice of edge labelling. 

The result of this matching and the subsequent solution of the linear
system is presented in the last column of
tables~\ref{tab:deltaresults} and~\ref{tab:epsresults}. This concludes
the computation of~\eqref{e:theintegral}.

\subsection{Covariant computation of the $R^4$ integral}

A useful check of our method is to compute the well-known~$R^4$ term
in the linearised heterotic or \mbox{type-IIB} theory, and verify that
it reproduces results which were obtained previously using
non-covariant methods.  The~$R^4$ term arises from a superfield
integral of the form
\begin{equation}
I_{R^4} = \int\!{\rm d}^{16}\theta\,\Phi^4\,,
\end{equation}
where the relevant terms in the scalar superfield~$\Phi$ take the form
\begin{equation}
\Phi = \ldots +
(\bar\theta\Gamma^{ijm}\theta)(\bar\theta\Gamma^{kl}{}_m\theta) R_{ijkl} + \ldots\,.
\end{equation}
By making explicit use of the results of the previous subsection, we
can express the integral in terms of the 24~independent parity-even
Lorentz singlets,
\begin{equation}
I_{R^4} = \sum_{i=1}^{24} \alpha_{i}\, T_{(i)}^{i_1j_1k_1\cdots i_8 j_8 k_8} \left[ \eta_{i_1i_2} R_{j_1k_1j_2k_2} \right] \cdots \left[ \eta_{i_7i_8} R_{j_7k_7j_8k_8} \right]\,.
\end{equation}
(Note that the parity-odd part of the integral does not contribute,
since there are no $\epsilon R^4$ scalars in ten dimensions.)  The
individual terms in the sum lead to lengthy linear combinations of the
26~independent quartic curvature scalars.  All dependence on the Ricci
curvature cancels in the total result, and by decomposing the result
on the basis of the 7~Fulling invariants~\cite{Fulling:1992vm} (see
also appendix~\ref{s:young}) we are left with the required result in
terms of Weyl tensors,
\begin{equation}
I_{R^4 } = 2^{24} 3^{4}\, \big(t_8
 t_8 + \tfrac{1}{8}\epsilon_{10}\epsilon_{10}\big) C^4
=
2^{32} 3^5 \left( -\tfrac{1}{4} C^{pqrs} C_{pq}{}^{tu}C_{rt}{}^{vw} C_{suvw} + C^{pqrs} C_p{}^t{}_r{}^u C_t{}^v{}_q{}^w C_{uvsw} \right) \,.
\label{e:r4intresult}
\end{equation}
Imposing this result actually turns out to be restrictive enough to
completely determine the coefficients $\alpha_1\dots \alpha_{24}$.

\section{Discussion and applications}
\label{s:discuss}

We have presented the computation of the ten-dimensional fermionic
integral~\eqref{e:theintegral} in terms of 33 basis tensors, using a
general method applicable to all fermionic integrals, in particular
those of ten- and eleven-dimensional supergravity. The solution of
this problem was facilitated by mapping it onto a graph construction
and colouring problem. To conclude, let us discuss a number of
applications of this integration procedure.

Let us start with the~$N=1$ heterotic theory. The use of the dilaton
superfield for the construction of the~$R^4$ invariant was proposed a
long time ago~\cite{nils2}. Although the expansion of the scalar
superfield was only worked out to lowest order, it should be possible
to extend this analysis to higher order in~$\theta$ using computer
assistance (a similar analysis would have to be performed to construct
the measure). One particular reason why it is interesting to pursue
this program is that it could provide insight into the still elusive
supersymmetrisation of the $(t_8 t_8 +
\frac{1}{8}\epsilon_{10}\epsilon_{10})R^4$ invariant. As was emphasised
in~\cite{Peeters:2000qj}, the $t_8 t_8$ part of this construction is
reasonably well understood because of its relation to a
super-Yang-Mills invariant. The $\epsilon\epsilon R^4$
term~\cite{dero3}, in contrast, does not admit a derivation by
``squaring'' a super-Yang-Mills action. However, this term potentially
plays an important role in the modification of the superspace torsion
constraints. A direct derivation in components would certainly help to
understand this issue.

The complete set of higher derivative interactions that accompany
the~$R^4$ term involve, in particular, fluxes that are essential for
understanding nontrivial compactifications of string/M-theory.
Unfortunately, there are intrinsic difficulties in the superspace
description of such interactions due to the absence of an off-shell
superspace formalism for theories with maximal supersymmetry.
However, a limited amount can be deduced from the available on-shell
superspace formulations.  For example, in the type-IIB theory the
superfield formulation of~\cite{howe1} encapsulates the classical
theory which contains component fields comprising the one-form (made
from a derivative of the complex scalar), the NS-NS and R-R
three-forms and the five-form field strength (as well as the
fermions). There are difficulties in extending this to the general
leading higher derivative interactions~\cite{deHaro:2002vk} but in the
special case in which only the five-form and the metric are
non-vanishing, the five-form dependence enters purely as a companion
to the curvature in the~$\theta^4$ term in the scalar
superfield~\cite{deHaro:2002vk,Green:2003an}. As shown
in~\cite{Green:2003an} this leads to an elegant understanding of the
nonrenormalisation of the $D3$-brane supergravity solution by these
leading higher derivative interactions.  In other examples string
perturbation theory has provided evidence for the structure of such
terms, but a superfield analysis has not yet been possible. For
example, certain~$H^2 R^3$ terms were determined from a string
calculation in~\cite{Peeters:2001ub,Frolov:2001xr}.  In any case, the
technical results of this paper may be of value in any future progress
towards an understanding of the superspace formulation of such
interactions.

A completely different application of our superintegration techniques
concerns the study of vertex operator correlators. In space-time
supersymmetric formalisms for superparticles, strings and membranes,
the leading higher-derivative amplitudes arise as simple expressions
over world-sheet fields, involving only a fermionic zero-mode
integral.  Such integrals are precisely of the type considered
here. Using the methods of the present paper, it has recently become
possible to determine the $(DF_{(4)})^2 R^2$ and $(DF_{(4)})^4$ terms
in the M-theory effective action directly from a superparticle
calculation~\cite{plef1}.

\section*{Acknowledgements}

We would like to thank Jos\'e Martin-Garcia, Jan Plefka and Steffen
Stern for discussions and correspondence. The work of CS was partially supported
by the U.S.~Department of Energy, grant no.~DE-FG02-03ER-41262.

\vfill\eject
\appendix
\section{Appendix: reduction of tensor polynomials}
\label{s:young}

Once one has done a superspace integral over a sufficiently complex
integrand, one typically ends up with tensor polynomials which can be
further reduced using the symmetries of the individual tensors, as
well as the exchange of identical tensors. We would here like to
comment briefly on a reduction method which incorporates \emph{all}
symmetries, including the Ricci cyclic identity and Bianchi identities
(this method was used implicitly in several papers, see
e.g.~\cite{Boulanger:2004zf,Peeters:2003pv}, but as far as we know was
never spelled out, and more importantly, does not seem to be widely
known; it is not mentioned in the standard
reference~\cite{Fulling:1992vm} and is also missed in most recent
literature on tensor polynomial simplification).

The simplest symmetries of tensors are \emph{mono-term} symmetries,
such as anti-symmetry in a set of indices. These symmetries always
relate one particular index distribution to one other distribution,
and can be used step-by-step to reduce a given tensor monomial to
a canonical form. Various efficient algorithms have been discussed in
the literature and we will not comment on these symmetries any further
(see~\cite{dres1,Portugal:1998qi,port2} and in particular the
implementation in~\cite{e_xact}).

The more complicated symmetries are \emph{multi-term} symmetries, such
as the Ricci cyclic identity or the Bianchi identity, which relate a
sum of terms with different index distributions. These symmetries are
all manifestations of the so-called \emph{Garnir} symmetries of Young
tableaux~\cite{b_saga1}. These state that the sum over all
anti-symmetrisation of boxes in a Garnir hook is identically
zero. Examples of such Garnir hooks are given below,
\begin{equation*}
\includegraphics*[width=.4\textwidth]{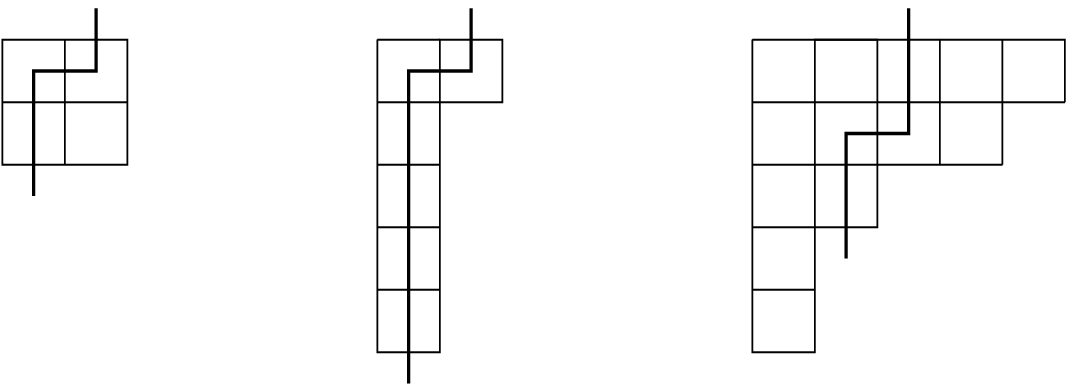}
\end{equation*}
which represent the Ricci cyclic identity, the Bianchi identity on a
five-form and a more general Garnir symmetry, respectively. Applying a
Garnir symmetry on a tensor produces a sum of tensors, which means
that one can no longer restrict to the canonicalisation of tensor
monomials.

There is, however, a simple way to construct instead a \emph{basis} of
monomials which takes the Garnir symmetries into account, and reduce
any given expression to this basis. It consists of simply replacing
each tensor in a monomial by its Young-projected form. A basis of
monomials can now be constructed by first generating a list of all
possible index contractions of the tensors, and then projecting each
of these using the Young projection on each of the individual
tensors. Monomials which are identical by virtue of Garnir symmetries
will then map to the same sum of monomials.

There are, however, a few subtleties. Firstly, it would be
prohibitively expensive to write down all terms in a Young-projected
tensor. Instead, it is much more efficient to reduce the
Young-projected forms by making use of the mono-term symmetries, which
are easy to deal with using the methods
of~\cite{dres1,Portugal:1998qi,port2}. One thus obtains, for e.g.~the
Riemann tensor,
\begin{equation}
\label{e:Rproj}
R_{a b c d} \rightarrow 
 \frac{1}{3}\big( 2\, R_{a b c d} - R_{a d b c} + R_{a c b d} \big)\,,
\end{equation}
instead of the $(2!)^4=16$ terms which are produced by the Young
projector.  The expression on the right-hand side manifestly satisfies
the cyclic Ricci identity, even if one only knows about the mono-term
symmetries of the Riemann tensor. Using the projector~\eqref{e:Rproj}
it is easy to show e.g.~that~$2\,R_{a b c d} R_{a c b d} = R_{a b c d}
R_{a b c d}$. The monomial on the left-hand side maps to
\begin{equation}
\begin{aligned}
R_{a b c d} R_{a c b d} \rightarrow \tfrac{1}{3} \big( R_{a b c d} R_{a c b d}
 + R_{a b c d} R_{a b c d} \big)\,,
\end{aligned}
\end{equation}
and similarly $R_{a b c d} R_{a b c d}$ maps to twice this expression,
thereby proving the identity in a way which easily extends to much
more complicated cases.

Secondly, the Young projectors are not dimension dependent. As a
result, not all relations between monomials will be recognised. The
additional relations exist because invariants sometimes arise as the
contraction of tensors with two $d$-dimensional epsilon tensors. When
written out in terms of Kronecker deltas, this leads to
anti-symmetrisation in~$d$ indices when none of the indices of the
epsilon tensors are contracted with each other. Clearly, such an
invariant can be written down in lower dimensions, but it vanishes
identically, although this is not recognised by simply performing the
Young projections. Relations obtained in this way have to be taken
into account separately, but are easy to find.

\setlength{\bibsep}{3pt}

\begingroup\raggedright\endgroup

\end{document}

%% file: proofzero11.pstex_t
\begin{picture}(0,0)%
\includegraphics{proofzero11.pstex}%
\end{picture}%
\setlength{\unitlength}{2960sp}%
\begingroup\makeatletter\ifx\SetFigFont\undefined%
\gdef\SetFigFont#1#2#3#4#5{%
  \reset@font\fontsize{#1}{#2pt}%
  \fontfamily{#3}\fontseries{#4}\fontshape{#5}%
  \selectfont}%
\fi\endgroup%
\begin{picture}(6322,780)(54,-151)
\put(1201,164){\makebox(0,0)[lb]{\smash{\SetFigFont{12}{14.4}{\rmdefault}{\mddefault}{\updefault}{\color[rgb]{0,0,0}$=$}%
}}}
\put(2701,164){\makebox(0,0)[lb]{\smash{\SetFigFont{12}{14.4}{\rmdefault}{\mddefault}{\updefault}{\color[rgb]{0,0,0}$=$}%
}}}
\put(3076,164){\makebox(0,0)[lb]{\smash{\SetFigFont{12}{14.4}{\rmdefault}{\mddefault}{\updefault}{\color[rgb]{0,0,0}$-$}%
}}}
\put(4651,164){\makebox(0,0)[lb]{\smash{\SetFigFont{12}{14.4}{\rmdefault}{\mddefault}{\updefault}{\color[rgb]{0,0,0}$\Rightarrow$}%
}}}
\put(6376,164){\makebox(0,0)[lb]{\smash{\SetFigFont{12}{14.4}{\rmdefault}{\mddefault}{\updefault}{\color[rgb]{0,0,0}$=0\,.$}%
}}}
\end{picture}

%% file: proofzero21.pstex_t
\begin{picture}(0,0)%
\includegraphics{proofzero21.pstex}%
\end{picture}%
\setlength{\unitlength}{2960sp}%
\begingroup\makeatletter\ifx\SetFigFont\undefined%
\gdef\SetFigFont#1#2#3#4#5{%
  \reset@font\fontsize{#1}{#2pt}%
  \fontfamily{#3}\fontseries{#4}\fontshape{#5}%
  \selectfont}%
\fi\endgroup%
\begin{picture}(4680,1312)(61,-533)
\put(1126, 14){\makebox(0,0)[lb]{\smash{\SetFigFont{12}{14.4}{\rmdefault}{\mddefault}{\updefault}{\color[rgb]{0,0,0}$=$}%
}}}
\put(2851, 14){\makebox(0,0)[lb]{\smash{\SetFigFont{12}{14.4}{\rmdefault}{\mddefault}{\updefault}{\color[rgb]{0,0,0}$=$}%
}}}
\put(3226, 14){\makebox(0,0)[lb]{\smash{\SetFigFont{12}{14.4}{\rmdefault}{\mddefault}{\updefault}{\color[rgb]{0,0,0}$(-)^3$}%
}}}
\end{picture}

%% file: graphs24.pstex_t
\begin{picture}(0,0)%
\includegraphics{graphs24.pstex}%
\end{picture}%
\setlength{\unitlength}{2960sp}%
\begingroup\makeatletter\ifx\SetFigFont\undefined%
\gdef\SetFigFont#1#2#3#4#5{%
  \reset@font\fontsize{#1}{#2pt}%
  \fontfamily{#3}\fontseries{#4}\fontshape{#5}%
  \selectfont}%
\fi\endgroup%
\begin{picture}(7597,7211)(1,-6383)
\put(5851,-3586){\makebox(0,0)[lb]{\smash{\SetFigFont{9}{10.8}{\rmdefault}{\mddefault}{\updefault}{\color[rgb]{0,0,0}16:}%
}}}
\put(  1,-4786){\makebox(0,0)[lb]{\smash{\SetFigFont{9}{10.8}{\rmdefault}{\mddefault}{\updefault}{\color[rgb]{0,0,0}17:}%
}}}
\put(6076,-4786){\makebox(0,0)[lb]{\smash{\SetFigFont{9}{10.8}{\rmdefault}{\mddefault}{\updefault}{\color[rgb]{0,0,0}20:}%
}}}
\put(3826,-4786){\makebox(0,0)[lb]{\smash{\SetFigFont{9}{10.8}{\rmdefault}{\mddefault}{\updefault}{\color[rgb]{0,0,0}19:}%
}}}
\put(3826,-3586){\makebox(0,0)[lb]{\smash{\SetFigFont{9}{10.8}{\rmdefault}{\mddefault}{\updefault}{\color[rgb]{0,0,0}15:}%
}}}
\put(3901,-5986){\makebox(0,0)[lb]{\smash{\SetFigFont{9}{10.8}{\rmdefault}{\mddefault}{\updefault}{\color[rgb]{0,0,0}23:}%
}}}
\put(2026,-3586){\makebox(0,0)[lb]{\smash{\SetFigFont{9}{10.8}{\rmdefault}{\mddefault}{\updefault}{\color[rgb]{0,0,0}14:}%
}}}
\put(2026,-4786){\makebox(0,0)[lb]{\smash{\SetFigFont{9}{10.8}{\rmdefault}{\mddefault}{\updefault}{\color[rgb]{0,0,0}18:}%
}}}
\put(2026,-5986){\makebox(0,0)[lb]{\smash{\SetFigFont{9}{10.8}{\rmdefault}{\mddefault}{\updefault}{\color[rgb]{0,0,0}22:}%
}}}
\put(  1,-5986){\makebox(0,0)[lb]{\smash{\SetFigFont{9}{10.8}{\rmdefault}{\mddefault}{\updefault}{\color[rgb]{0,0,0}21:}%
}}}
\put(1051,-2461){\makebox(0,0)[lb]{\smash{\SetFigFont{9}{10.8}{\rmdefault}{\mddefault}{\updefault}{\color[rgb]{0,0,0}$\times$}%
}}}
\put(  1,-2461){\makebox(0,0)[lb]{\smash{\SetFigFont{9}{10.8}{\rmdefault}{\mddefault}{\updefault}{\color[rgb]{0,0,0}10:}%
}}}
\put(676,464){\makebox(0,0)[lb]{\smash{\SetFigFont{9}{10.8}{\rmdefault}{\mddefault}{\updefault}{\color[rgb]{0,0,0}$\times$}%
}}}
\put(1126,464){\makebox(0,0)[lb]{\smash{\SetFigFont{9}{10.8}{\rmdefault}{\mddefault}{\updefault}{\color[rgb]{0,0,0}$\times$}%
}}}
\put(1576,464){\makebox(0,0)[lb]{\smash{\SetFigFont{9}{10.8}{\rmdefault}{\mddefault}{\updefault}{\color[rgb]{0,0,0}$\times$}%
}}}
\put(  1,464){\makebox(0,0)[lb]{\smash{\SetFigFont{9}{10.8}{\rmdefault}{\mddefault}{\updefault}{\color[rgb]{0,0,0}1:}%
}}}
\put(6301,464){\makebox(0,0)[lb]{\smash{\SetFigFont{9}{10.8}{\rmdefault}{\mddefault}{\updefault}{\color[rgb]{0,0,0}$\times$}%
}}}
\put(6751,464){\makebox(0,0)[lb]{\smash{\SetFigFont{9}{10.8}{\rmdefault}{\mddefault}{\updefault}{\color[rgb]{0,0,0}$\times$}%
}}}
\put(5626,464){\makebox(0,0)[lb]{\smash{\SetFigFont{9}{10.8}{\rmdefault}{\mddefault}{\updefault}{\color[rgb]{0,0,0}3:}%
}}}
\put(2926,464){\makebox(0,0)[lb]{\smash{\SetFigFont{9}{10.8}{\rmdefault}{\mddefault}{\updefault}{\color[rgb]{0,0,0}2:}%
}}}
\put(3601,464){\makebox(0,0)[lb]{\smash{\SetFigFont{9}{10.8}{\rmdefault}{\mddefault}{\updefault}{\color[rgb]{0,0,0}$\times$}%
}}}
\put(4051,464){\makebox(0,0)[lb]{\smash{\SetFigFont{9}{10.8}{\rmdefault}{\mddefault}{\updefault}{\color[rgb]{0,0,0}$\times$}%
}}}
\put(6376,-511){\makebox(0,0)[lb]{\smash{\SetFigFont{9}{10.8}{\rmdefault}{\mddefault}{\updefault}{\color[rgb]{0,0,0}$\times$}%
}}}
\put(5701,-511){\makebox(0,0)[lb]{\smash{\SetFigFont{9}{10.8}{\rmdefault}{\mddefault}{\updefault}{\color[rgb]{0,0,0}6:}%
}}}
\put(5776,-1486){\makebox(0,0)[lb]{\smash{\SetFigFont{9}{10.8}{\rmdefault}{\mddefault}{\updefault}{\color[rgb]{0,0,0}9:}%
}}}
\put(6751,-1486){\makebox(0,0)[lb]{\smash{\SetFigFont{9}{10.8}{\rmdefault}{\mddefault}{\updefault}{\color[rgb]{0,0,0}$\times$}%
}}}
\put(3676,-511){\makebox(0,0)[lb]{\smash{\SetFigFont{9}{10.8}{\rmdefault}{\mddefault}{\updefault}{\color[rgb]{0,0,0}$\times$}%
}}}
\put(2926,-511){\makebox(0,0)[lb]{\smash{\SetFigFont{9}{10.8}{\rmdefault}{\mddefault}{\updefault}{\color[rgb]{0,0,0}5:}%
}}}
\put(2926,-1486){\makebox(0,0)[lb]{\smash{\SetFigFont{9}{10.8}{\rmdefault}{\mddefault}{\updefault}{\color[rgb]{0,0,0}8:}%
}}}
\put(3901,-1486){\makebox(0,0)[lb]{\smash{\SetFigFont{9}{10.8}{\rmdefault}{\mddefault}{\updefault}{\color[rgb]{0,0,0}$\times$}%
}}}
\put(676,-511){\makebox(0,0)[lb]{\smash{\SetFigFont{9}{10.8}{\rmdefault}{\mddefault}{\updefault}{\color[rgb]{0,0,0}$\times$}%
}}}
\put(  1,-511){\makebox(0,0)[lb]{\smash{\SetFigFont{9}{10.8}{\rmdefault}{\mddefault}{\updefault}{\color[rgb]{0,0,0}4:}%
}}}
\put(676,-1486){\makebox(0,0)[lb]{\smash{\SetFigFont{9}{10.8}{\rmdefault}{\mddefault}{\updefault}{\color[rgb]{0,0,0}$\times$}%
}}}
\put(  1,-1486){\makebox(0,0)[lb]{\smash{\SetFigFont{9}{10.8}{\rmdefault}{\mddefault}{\updefault}{\color[rgb]{0,0,0}7:}%
}}}
\put(2926,-2461){\makebox(0,0)[lb]{\smash{\SetFigFont{9}{10.8}{\rmdefault}{\mddefault}{\updefault}{\color[rgb]{0,0,0}11:}%
}}}
\put(6001,-2461){\makebox(0,0)[lb]{\smash{\SetFigFont{9}{10.8}{\rmdefault}{\mddefault}{\updefault}{\color[rgb]{0,0,0}12:}%
}}}
\put(  1,-3586){\makebox(0,0)[lb]{\smash{\SetFigFont{9}{10.8}{\rmdefault}{\mddefault}{\updefault}{\color[rgb]{0,0,0}13:}%
}}}
\put(6226,-5986){\makebox(0,0)[lb]{\smash{\SetFigFont{9}{10.8}{\rmdefault}{\mddefault}{\updefault}{\color[rgb]{0,0,0}24:}%
}}}
\end{picture}

%% file: zerographs.pstex_t
\begin{picture}(0,0)%
\includegraphics{zerographs.pstex}%
\end{picture}%
\setlength{\unitlength}{2960sp}%
\begingroup\makeatletter\ifx\SetFigFont\undefined%
\gdef\SetFigFont#1#2#3#4#5{%
  \reset@font\fontsize{#1}{#2pt}%
  \fontfamily{#3}\fontseries{#4}\fontshape{#5}%
  \selectfont}%
\fi\endgroup%
\begin{picture}(6580,4286)(461,-3608)
\end{picture}

%% file: epsgraphs.pstex_t
\begin{picture}(0,0)%
\includegraphics{epsgraphs.pstex}%
\end{picture}%
\setlength{\unitlength}{2960sp}%
\begingroup\makeatletter\ifx\SetFigFont\undefined%
\gdef\SetFigFont#1#2#3#4#5{%
  \reset@font\fontsize{#1}{#2pt}%
  \fontfamily{#3}\fontseries{#4}\fontshape{#5}%
  \selectfont}%
\fi\endgroup%
\begin{picture}(7319,4378)(22,-3560)
\put(6901,-1036){\makebox(0,0)[lb]{\smash{\SetFigFont{9}{10.8}{\rmdefault}{\mddefault}{\updefault}{\color[rgb]{0,0,0}$5$}%
}}}
\put(826,-2686){\makebox(0,0)[lb]{\smash{\SetFigFont{9}{10.8}{\rmdefault}{\mddefault}{\updefault}{\color[rgb]{0,0,0}$\times$}%
}}}
\put(826,-136){\makebox(0,0)[lb]{\smash{\SetFigFont{9}{10.8}{\rmdefault}{\mddefault}{\updefault}{\color[rgb]{0,0,0}$\times$}%
}}}
\put(5401,-1036){\makebox(0,0)[lb]{\smash{\SetFigFont{9}{10.8}{\rmdefault}{\mddefault}{\updefault}{\color[rgb]{0,0,0}$4$}%
}}}
\put(3826,-1036){\makebox(0,0)[lb]{\smash{\SetFigFont{9}{10.8}{\rmdefault}{\mddefault}{\updefault}{\color[rgb]{0,0,0}$3$}%
}}}
\put(2251,-1036){\makebox(0,0)[lb]{\smash{\SetFigFont{9}{10.8}{\rmdefault}{\mddefault}{\updefault}{\color[rgb]{0,0,0}$2$}%
}}}
\put(4726,-3511){\makebox(0,0)[lb]{\smash{\SetFigFont{9}{10.8}{\rmdefault}{\mddefault}{\updefault}{\color[rgb]{0,0,0}$8$}%
}}}
\put(2551,-3511){\makebox(0,0)[lb]{\smash{\SetFigFont{9}{10.8}{\rmdefault}{\mddefault}{\updefault}{\color[rgb]{0,0,0}$7$}%
}}}
\put(451,-3511){\makebox(0,0)[lb]{\smash{\SetFigFont{9}{10.8}{\rmdefault}{\mddefault}{\updefault}{\color[rgb]{0,0,0}$6$}%
}}}
\put(6826,-3511){\makebox(0,0)[lb]{\smash{\SetFigFont{9}{10.8}{\rmdefault}{\mddefault}{\updefault}{\color[rgb]{0,0,0}$9$}%
}}}
\put(451,-1036){\makebox(0,0)[lb]{\smash{\SetFigFont{9}{10.8}{\rmdefault}{\mddefault}{\updefault}{\color[rgb]{0,0,0}$1$}%
}}}
\end{picture}